\documentclass[11pt]{article}

\usepackage{fleqn,cospar}

\usepackage{url}

\usepackage{graphicx}
\usepackage[figuresright]{rotating}

\title{SPECTRAL PROPERTIES OF A SAMPLE OF LOW LUMINOSITY TYPE~I X-RAY BURSTERS}

\author{L. Natalucci}

\begin{document}

\maketitle
\vspace{0.2cm}
\centerline{\it{Istituto di Astrofisica Spaziale (IAS/CNR),
    Via Fosso del Cavaliere 100,00133 Roma, Italy}}
\vspace{0.2cm} 
\centerline{on behalf of the {\em BeppoSAX}/WFC Galactic Bulge collaboration}

\begin{abstract}
I report on a sample of new type~I X-ray bursters, firstly detected 
with the Wide Field Cameras on board {\em BeppoSAX} and then studied with the
Narrow Field Instruments on a broad spectral range (0.1-200 keV).
Properties of the transient/persistent emission are summarized and the
broad band X-ray spectra discussed in detail for a few sources. 

\end{abstract}

\section*{INTRODUCTION}

Short transient event phenomena like X-ray bursts are known since 1975 
(Grindlay et al. 1976, Belian et al. 1976), with a total of $\sim20$ 
burst sources revealed within about one year since the initial 
discovery. 
Accretion onto the surface of a weakly magnetized neutron star (NS) 
can give rise to X-ray bursts via unstable
nuclear burning, if the system is filled through Roche-lobe overflow at 
transfer rates much lower than the Eddington limit of 
$\sim2\times10^{-8}$~${M}_{\odot}$/year (van Paradijs et al.,1988).  
Wind-fed systems cannot be X-ray burst sources whereas 
most persistent X-ray bursters are identified with low mass X-ray binaries 
(LMXB), being mostly 
Atoll sources. Relevant source properties can be derived from combined 
burst and persistent emission study (distance, radius, composition of 
nuclear fuel etc., see Lewin et al. 1995 for review). In this respect, 
very important are the so-called {\it super-Eddington} bursts, in which
expansion of the blackbody (BB) emitting photosphere occurs up to radii of 
several tenths of kilometers (Lewin et al. 1993), due to the emission 
level reaching the Eddington luminosity.
The profiles from these bursts tend to flatten at energies above 
$\sim5$~keV and often show a double-peaked structure, each peak  
being associated with the expansion and contraction phases (see Figure~1).
This allows to estimate distances via the determination of the burst peak 
fluxes, assumed to be at a level of $\sim2\times10^{38}$~erg~s$^{-1}$ 
which is the Eddington luminosity of a standard 1.4~${M}_{\odot}$~NS.

Recently, X-ray burst research went to a new exciting phase following the
the discovery of high time resolution features like kHz nearly coherent 
oscillations in the burst time profiles (Strohmayer et al. 1996), 
present during both the rising and cooling phases and then probably 
associated to the NS spin period (Strohmayer \& Markwardt, 1999). These 
studies can also give important insight in the understanding of the geometry, 
structure and composition of the burning layers (Muno et al., 2000). 
 
Since 1992, the detection by GRANAT/SIGMA of hard X-ray emission from 
a few bursters, namely KS1731-260, GX354+0 and 1E1724-3045 
(Barret \& Vedrenne 1994) and subsequently other BATSE
observations (Barret et al. 1996) have established X-ray bursters as a new 
important class of low luminosity hard X-ray emitters. A few years after 
these early '90 observations, the population of X-ray bursters had a  
substantial increase by means of the {\em BeppoSAX} observation campaigns 
performed since 1996 with the Wide Field Cameras (WFC). 
These allowed to discover 16 new bursters thus increasing the known 
population by about 30\%.
Among these objects, interesting is the discovery of burst emission from 
GS~1826-238 (Ubertini et al. 1999a) showing that this source, previously
considered as a transient Black Hole Candidate (BHC) is instead a 
persistent, low luminosity source with nearly periodic, extremely regular 
bursting behaviour.  

In this work I report about observations of a sample of these new bursters
and discuss their broad band spectral properties, as obtained from
Target-Of-Opportunity observations by the Narrow Field Instruments (NFI) on 
board {\em BeppoSAX}. 
This is a set of co-aligned telescopes capable of efficient, continuos 
spectral coverage in the band 0.1-220~keV (Boella et al., 1997).

\section*{THE SOURCES}

Burst detection is the objective tool to distinguish between BH and 
NS binaries. In case of NS LMXB transients with sporadic X-ray outbursts,
type~I bursts are often seen when the persistent emission level is above a 
few percent of the Eddington limit, whereas bursting activity is generally 
absent in quiescent state. Long term monitoring is then essential 
not also for discovering of new transients, but also for their 
characterization as NS or BH binaries.

Since LMXBs are highly concentrated towards the Galactic Bulge, a 
sensitive monitoring of this region can be obtained with pointed observations 
of wide field instruments. For this purpose, a monitoring 
program is being performed since August 1996 by the {\em BeppoSAX}/WFC 
(Jager et al 1997) by means of observations repeated periodically during 
the Spring and Fall of each year. The monitoring itself has a 
sensitivity of a few mCrab, that is an order of magnitude beyond the 
capabilities of current All-Sky monitors. At the distance of the Galactic 
Centre this sensitivity limit corresponds to a 2-10~keV source flux of 
less than $10^{36}$~erg~s$^{-1}$. This has allowed the discovery of a new 
population of low luminosity NS transients in the Galactic Bulge, probably 
consisting of short period binaries with evolved mass companions 
(Heise et al. 1999). 
Binary systems harboring weakly magnetized NS are known to be preferentially
persistent. This is possibly related to system thermal stability being 
induced by X-ray irradiation of the outer disk (e.g., King et al. 1997). This 
condition is more easily found in NS than in black hole binaries,
but indeed it may be expected that NS systems with very low accretion rates
can be transient (King 2000).  
Currently, 9 new faint transients have been discovered by the 
{\em BeppoSAX}/WFC. Most of them have been observed bursting, namely: 
the 2.5~ms accreting pulsar SAX~J1808.4-3658 (in't Zand et al. 1998a), 
SAX~J1748.9-2021 in the globular cluster NGC~6440 (in't Zand et al. 1999a), 
SAX~J1750.8-2900 (Natalucci et al. 1999), SAX~J1712.6-3739 
(Cocchi et al. 1999a) and SAX~J1810.8-2609 (Natalucci et al. 2000a).  
Burst events were also observed from three other sources, for which no 
persistent emission was detected: SAX~J1753.5-2349, SAX~J1806.5-2215 
(in't Zand et al. 1998b) and SAX~J1752.3-3138 (Cocchi et al. 1999b).
From another new faint transient, SAX~J1819.2-2525 (in't Zand et al. 2000) 
no bursts were detected so far.

The fact that the majority of these weak transients are NS is reminiscent of 
the observed behavior in persistent sources, where it is found that low state 
X-ray bursters are much less luminous than the corresponding low state BH 
sources (typically by a factor 10 to 100, see e.g. Barret et al. 2000).

For this work, I have selected a sample of X-ray bursters for which broad band 
spectra were obtained
by means of a related {\em BeppoSAX} TOO program (see Table I) or by 
dedicated observations. This sample contains two weak transients, one 
recurrent transient (SAX~J1747.0-2853)
and two persistent sources (GS~1826-238 and SLX 1735-269). In 
remainder of the paper, I will outline the most relevant spectral
properties of these sources.

\begin{table}[h]
\vspace{-8mm}
\begin{minipage}{180mm}
  \caption{The selected sample of X-ray bursters observations performed by the {\sl BeppoSAX}/NFI}
\begin{tabular}{lcccc}
\hline
Source Name  &  Obs. Date & Duration  & $F_{2-10 keV}$  & References \\
             &  & (ksec)  & ($10^{-10}$ erg~s$^{-1}$~cm$^{-2}$)  &\\
\hline
SAX J1810.8-2609  & 1998 Mar 12-13  & 85 & 4.3 & Natalucci et al. (2000a)\\
SAX J1712.6-3739  & 1999 Aug 27     & 45 & 1.2 & Cocchi et al. (1999a)\\
SAX J1747.0-2853  & 1998 Mar 23-24  & 72 & 3.0 & Natalucci et al. (2000b)\\
GS 1826-238       & 1997 Apr 6-7    & 41 & 5.4 & in't Zand et al. (1999b) \\
                  & 1999 Oct 20-21  & 60 & 6.3 & Cocchi et al., in prep. \\
SLX 1735-269      & 1997 Sep 18-19  & 61 & 5.0 & Bazzano et al., in prep. \\
\hline
\end{tabular}
\end{minipage}
\end{table}

\subsection*{SAX J1810.8-2609: a faint transient with a very hard spectrum}

This source was discovered by the {\em BeppoSAX}/WFC on 1998 
March 10 (Ubertini et al., 1998a), with a persistent emission level of 
$3.1\times10^{-10}$ erg~s$^{-1}$~cm$^{-2}$ ($\sim15$~mCrab) in the 
2-10~keV band. The source position is consistent with that of a soft
X-ray source, RX~J1810.7-2609, revealed by ROSAT during a follow-up
observation performed 15 days later (Greiner et al., 1998).  
On March 11, i.e. one day after its discovery a strong, super-Eddington 
type~I burst was detected. This allowed an estimate of the source distance   
of $\sim5$~kpc. The burst has the characteristic radius expansion profile
(Figure 1) and a very fast $\sim1$~sec) rise time, typical of a helium 
burning flash (Lewin et al.1993). 

SAX J1810.8-2609 was observed again by {\em BeppoSAX} about two days after 
its discovery, using the NFI. 
The X-ray spectrum from the source is exceptionally hard for a NS system, 
as the slope of the high energy power law is 1.96 and there is no visible 
high energy cutoff (Figure 2). The best fit spectrum, yielding 
$\chi^2_\nu$=0.99 (161 dof)
is obtained as a sum of two components: a blackbody spectrum with a
$\sim0.4$~keV temperature, and a hard tail which results from Compton
up-scattering of soft seed photons, represented by a thermal distribution 
at $\sim0.6$~keV (see the referenced paper for details). The presence of
the soft blackbody component, with a total flux of 
$\sim3\times10^{-11}$ erg~s$^{-1}$~cm$^{-2}$ is highly significant, as
the estimated chance probability evaluated via F-test is 
$\sim7\times10^{-9}$ for the model given above. For this source, a 
two-component blackbody plus power law 
gives also a good fit ($\chi^2_\nu$=0.97 for 163 dof) but only the former
model, which involves thermal comptonization is capable of matching the 
estimated value of Galactic column absorption of $3.7\times10^{21}$~cm$^{-2}$. 

\begin{figure}
\vspace{-0.5cm}
\begin{minipage}[b!]{7.5cm}
\includegraphics[width=7.5cm]{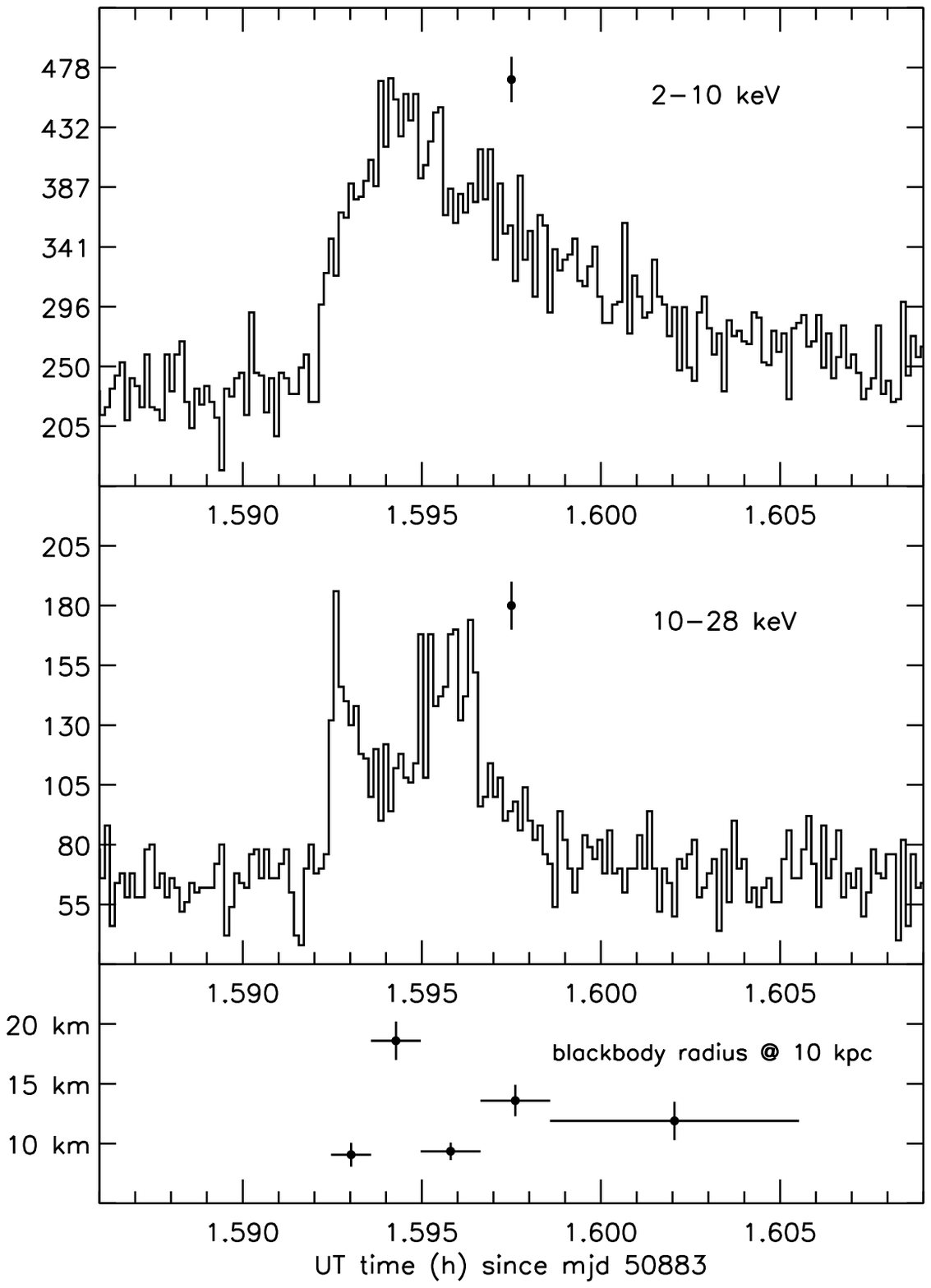}
\vspace{-1.2cm}
\caption{Time profiles of the March 11.06634 type~I X-ray burst from
the weak transient SAX~J1810.8-2609. The ratio blackbody
radius vs. distance is shown in units of km/10~kpc.}
\vspace{0.5cm}
\label{fig1}
\end{minipage}%
\hspace*{1.5cm}
\begin{minipage}[b!]{9.0cm}
\vspace*{1.8cm}
\vspace{-0.8cm}
\includegraphics[width=9cm]{Fig2.ps}
\vspace{-1.2cm}
\caption{The broadband spectrum of SAX~J1810.8-2609 measured by the NFI
on 1998 March 12-13. The model fit is as explained in text.}
\end{minipage}
\end{figure}

\subsection*{SAX J1712.6-3739}
This new, faint transient object was detected for the first time 
on 1999, August 25 by the WFC (in 't Zand et al., 1999c), at a position 
$\sim0.6$~arcmin from a ROSAT all-sky-survey source, 1RXS~J171237.1-373834.
The source was designated as a candidate NS LMXB, following the WFC detection 
of a long (lasting $\sim20$~s) type~I X-ray burst on September 2
(Cocchi et al. 1999a) having a peak intensity of 1.7 Crab. On August 27 the 
NFI were pointed on SAX~J1712.6-3739 for a TOO observation, when the source 
intensity was $\sim6$~mCrab in the 2-10 keV band. The measured spectrum is 
quite hard and is well fitted by unsaturated thermal comptonization with 
seed photons temperature of $0.47\pm0.03$~keV, and the resulting $\chi^2_\nu$
is 1.00 for 117 dof. The overall flux, however, was too weak to determine
the cutoff energy of the comptonized tail. 

This weak outburst and the related type~I X-ray burst emission were the only 
observed events from SAX~J1712.6-3739 throughout about five years of 
monitoring.  

\subsection*{The recurrent transient SAX J1747.0-2853 very close to the 
Galactic Centre}

SAX J1747.0-2853 is a rather intriguing source. Until last year it was 
known to have recurrent weak outbursts, i.e. at a level less than $\sim20$
mCrab. Since its discovery in 1998 by in't Zand et al. (1998c), 
two faint outburst events were observed by the 
WFC in spring of 1998 and spring of 1999. In March 2000 the source 
re-appeared with an average intensity of 
$\approx42$~mCrab in the 2-10~keV band, increasing up to 140 mCrab on 
March~7.9 as observed by the {\em RXTE}~PCA (Markwardt et al. 2000) and 
{\em BeppoSAX} (Campana et al. 2000). 

\begin{figure}
\begin{minipage}[b!]{9cm}
\vspace{1.0cm}
\includegraphics[width=9cm]{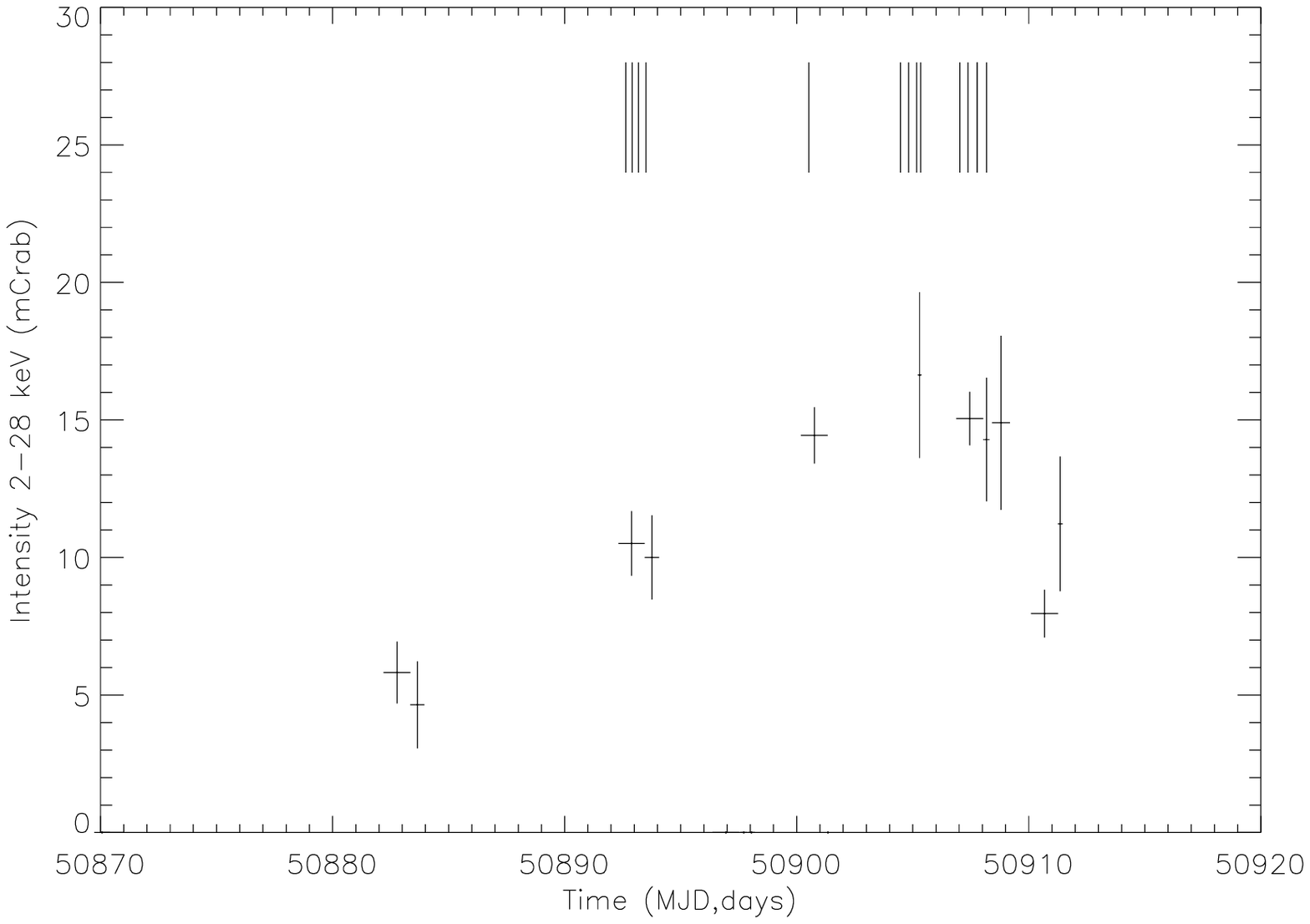}
\vspace{-1.0cm}
\caption{Light curve of the 1998 outburst from SAX~J1747.0-2853 in the
2-28~keV band. Burst detection times are indicated by markers.}
\vspace{0.5cm}
\label{fig3}
\end{minipage}%
\hspace*{1.2cm}
\begin{minipage}[b!]{7.5cm}
\includegraphics[width=7.5cm]{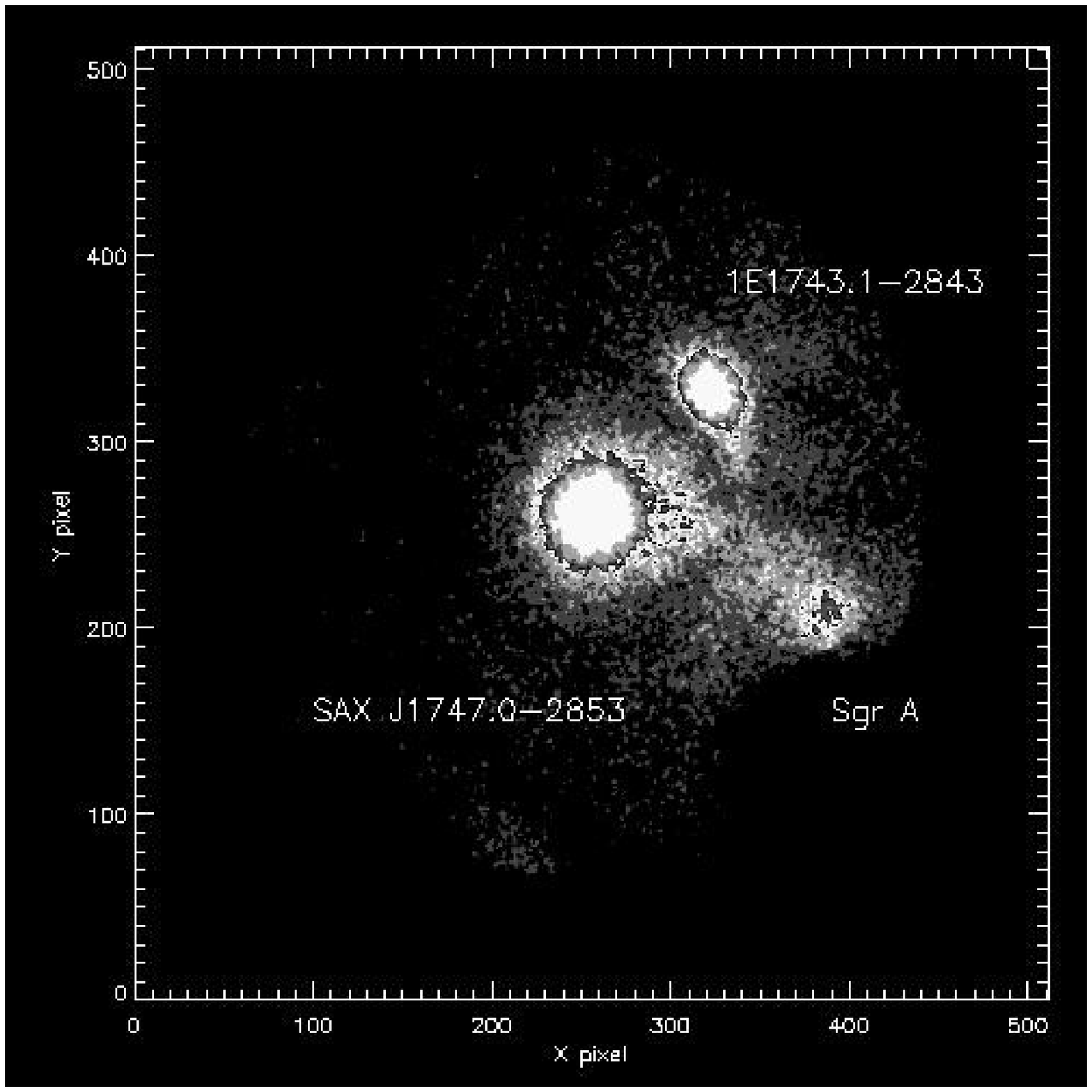}
\vspace{-1.2cm}
\caption{{\em BeppoSAX}/MECS image of the 1 square degree field around 
SAX~J1747.0-2853, taken on 1998 March 23-24.}
\vspace{0.5cm}
\label{fig4}
\end{minipage}%
\end{figure}

A total of 14 X-ray bursts were detected during the 1998 Spring campaign. 
One of the strongest events clearly showed radius expansion and allowed to 
estimate a distance of $\sim9$~kpc (Natalucci et al., 2000b). The outburst
X-ray light curve has a slowly rising edge, with intensity 
increasing at a rate of $\sim1.1\times10^{35}$~erg~s$^{-1}$~day$^{-1}$ 
(see Figure~3). The peak luminosity for this outburst is 
$\sim3\times10^{36}$~erg~s$^{-1}$ in the 2-10 keV band. The source was 
pointed with the NFI in March 23-24. Figure~4 shows the MECS image, obtained
in the 0.5-10.5 keV band, of the $\sim1$~square degree region centered on the 
object. Two other objects are present in the field, one is identified
with the SgrA complex and the other is a soft spectrum source, 
1E~1743.1-2843. Up to 10.5~keV, the presence of additional sources is not 
a problem when careful background subtraction is performed. At higher 
energies, however, the lack of spatial resolution of the {\em BeppoSAX}
collimated instruments leads to source confusion. 

Under reasonable assumptions and excluding the spectral channels between 
10.5 and 32 keV (see Natalucci et al., 2000b for all details on data 
selection), 
a broadband spectrum could be obtained as shown in Figure~4. The primary 
emission from SAX~J1747.0-2853 can be described by a 
combination of thermal comptonization plus a soft blackbody component at
$\sim0.55\pm0.10$~keV. A sharp iron absorption edge is detected at 
an energy of $\sim7.4\pm0.2$~keV (see Figure 6), together with a narrow 
K$\alpha$ line at $6.9\pm0.1$~keV, with an equivalent width of $\sim55$~eV. 
The significance of the line is high (the F-test gives a $1.4\times10^{-5}$   
chance probability).
However, as a thermal plasma with line emission is present in
the Galactic Centre region (Koyama et al., 1996), systematics could be 
induced in the 
background subtraction process by the non-uniform spatial distribution 
of this diffuse emission. For this reason this detection should be not 
considered a straight evidence.

\begin{figure}
\vspace{-0.5cm}
\begin{minipage}[b!]{9cm}
\vspace*{0.5cm}
\includegraphics[width=9cm]{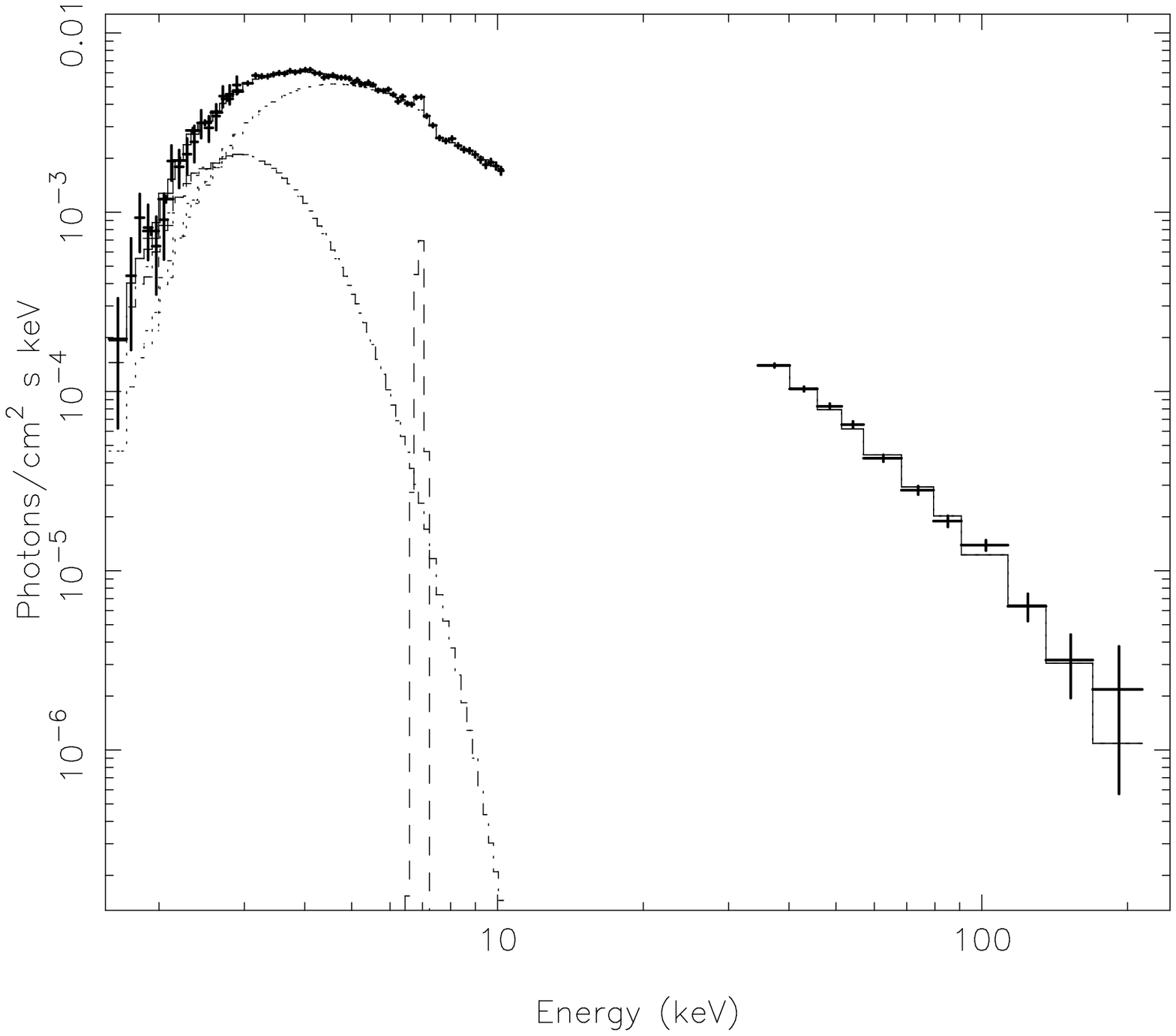}
\vspace{-0.5cm}
\caption{The hard/low state spectrum of SAX~J1747.0-2853 detected by the 
NFI in spring of 1998.}
\label{fig5}
\end{minipage}%
\hspace*{1.2cm}
\begin{minipage}[b!]{7.5cm}
\vspace{0.9cm}
\includegraphics[width=6.5cm]{Fig6.ps}
\caption{The SAX~J1747.0-2853 count rate spectrum and residuals to a 
power law fit are shown in the range 1.8-10.5 keV. A strong, sharp
absorption edge is present at $\sim7.4$~keV.}
\vspace{-0.8cm}
\label{fig6}
\end{minipage}
\end{figure}

\subsection*{GS~1826-238}
GS~1826-238 flux measurements throughout several years established 
this X-ray binary to be a faint, persistent source. The WFC discovery of
extremely regular bursting intervals, with a slowly evolving 
quasi-periodicity (Ubertini et al. 1999a,1999b) 
has recently raised great interest around this source, formerly believed 
to be a BHC. An up-to-date description of the long term bursting 
behaviour can be found in Cocchi et al (2001). 

The source was observed several times by the {\em BeppoSAX}/NFI. Del Sordo
et al. (1999) fitted a broad band spectrum with a blackbody component and
high energy tail represented by a cutoff power law. A two component model
was also suggested by int'Zand et al. (1999b), in which the hard 
X-ray emission is accounted by unsaturated thermal comptonization 
(Titarchuk, 1994; Hua \& Titarchuk, 1995). Using the same model, 
good quality fits ($\chi^2_\nu$~$\sim1.1$) have been obtained for an 
observation performed during the Fall of 1999. In this case the spectrum is 
compatible with the sum of a comptonized emission with seed photon 
temperature $\sim1.3$~keV and a soft blackbody component at 
$\sim0.6$~keV (Cocchi et al., in preparation). 
For the latter, the ratio BB radius/distance is 
$\sim100$~km/kpc, hence compatible with a typical NS radius.

\subsection*{SLX1735-269}
This persistent source, recently discovered as a new burster 
(Bazzano et al., 1997) was observed by the NFI on Fall 1999, showing
a persistent flux of 
$\sim5\times10^{-10}$ erg~s$^{-1}$~cm$^{-2}$ in the 0.5-150~keV band. From
a preliminary analysis, the 
spectrum is fitted by a single component model consisting
of unsaturated thermal comptonization with seed photon temperature of
$0.48\pm0.02$~keV. There is weak indication of a high energy cutoff,
corresponding to ${kT}_{e}$=$26\pm11$~keV. This fit yields 
$\chi^2_\nu$=1.15 for 101 dof, and an absorption coefficient 
${N}_{H}$=$0.74\pm0.08$~cm$^{-2}$. No additional blackbody component is
detected.

\section*{DISCUSSION}
The sample of X-ray bursters considered has remarkable common
properties. The luminosities found are generally low, below 
$\sim10^{-37}$ erg~s$^{-1}$~cm$^{-2}$, with detection of hard tails in
all the sources. The analysis of broadband spectra reveals unsaturated
thermal comptonization as the primary mechanism for the hard X-ray source
in these X-ray bursters. This is also confirmed by other observations of 
similar NS (Guainazzi et al., 1999). An additional soft component is 
observed in SAX~J1810.8-2609, SAX~J1747.0-2853 and GS~1826-238, well 
described by blackbody emission at
a temperature of $\sim0.5$~keV. In SLX1735-238 and SAX~J1712.6-3739 this
soft component is not necessary to obtain a good fit, possibly due to 
the lower statistics. 

The analysis of the spectra reveal a preference for two different 
sources of thermal photons, corresponding to the blackbody (apparently
unscattered) and to the comptonized seed photons. A systematic study is
on going to determine the relevant properties of these two components,
in particular to constrain the origin of the emission. For the
sources observed, the blackbody component is generally compatible
with a photosphere radius of the order of (or slightly greater than) 
10~km, so indicative of a NS origin. If the soft component is fitted with
a multicolor disk model, the values of the inner radii are those
expected from an inner disk quite close to the NS ($\sim10$~km to several
tenths of km). Detections of blackbody temperatures around $\sim2$~keV
are sometimes reported for low/hard states (e.g., Balucinska-Church et 
al. 1999, Barret et al. 2000), but in this case the 
normalization often yields blackbody radii much smaller than the
NS radius (typically less than $\sim1$~km). Incidentally, this happens 
mostly when a cutoff power law is used to model the primary hard
component instead of thermal comptonization with a Wien-like input 
spectrum. 

The situation is more complex for what concerns the comptonized 
component. For SAX~J1748.9-2021 in't~Zand et al. (1999a) report 
a seed temperature of 0.57~keV which is compatible with a radius very 
close to the NS, $\sim13$~km. Values close to $\sim0.5$~keV are also 
obtained from SLX~1735-269 and SAX~J1810.8-2609. For the former source,
this is consistent with RXTE data, with the report of a $0.53\pm0.07$~keV
temperature (Barret et al., 2000). In other sources 
(GS~1826-238, SAX~J1747.0-2843) temperatures in the range 1-1.2~keV
are reported.  For the electron plasma, there is  
a "clustering" of ${kT}_{e}$ values around $\sim20$~keV, apart from 
SAX~J1810.8-2609 which shows a rather peculiar and very hard tail (no
cutoff is observed). Its spectrum is reminiscent of that of a low state BHC 
where the high energy cutoff is located above $\sim100$~keV. 

\section*{ACKNOWLEDGEMENTS}

The author would like to thank the WFC Galactic Bulge Collaboration, and
in particular J.Heise, P.Ubertini, A.Bazzano, M.Cocchi, J.J.M in't Zand and 
E.Kuulkers for providing many results outlined in this paper. I am also 
grateful to Team Members of the BeppoSax Science Operation Centre and 
Science Data Centre for the continuos support. The {\em BeppoSAX} satellite 
is a joint Italian and Dutch programme.

\end{document}